\documentclass[jgrga]{agu2001}
\usepackage{graphicx}

\authorrunninghead{SAGAR ET AL.}

\titlerunninghead{AEROSOL PROPERTIES OVER MANORA PEAK}

\authoraddr{Ram Sagar, Brijesh Kumar, U.C. Dumka and P. Pant, State Observatory, Manora Peak, Nainital, Uttaranchal, 
            India.  (sagar@upso.ernet.in, brij@upso.ernet.in, dumka@upso.ernet.in, ppant@upso.ernet.in)}
\authoraddr{K. Krishna Moorthy, Space Physics Laboratory, Vikram Sarabhai Space Centre, Trivandrum, India. (k-k-moorthy@eth.net)}
             
\begin{document}

\title{Characteristics of Aerosol Spectral Optical Depths over Manora Peak $-$ A High Altitude Station in the Central 
       Himalayas \thanks{Tables 2, 3 and 4 are available in electronic form and can be obtained from the JGR electronic 
       database or from authors.}}

\authors{Ram Sagar,\altaffilmark{1} Brijesh Kumar,\altaffilmark{1} U.C. Dumka,\altaffilmark{1}
         K. Krishna Moorthy, \altaffilmark{2} and P. Pant \altaffilmark{1}}

\altaffiltext{1}
 {State Observatory, Manora Peak, Nainital, Uttaranchal, India}

\altaffiltext{2}
 {Space Physics Laboratory, Vikram Sarabhai Space Centre, Trivandrum, India }

\begin{abstract}
    We present, for the first time, spectral behaviour of aerosol optical depths (AODs) over Manora Peak, Nainital
    located at an altitude of $\sim$ 2 km in the Shivalik ranges of central Himalayas. The observations were carried out using a
    Multi-Wavelength solar Radiometer during January to December 2002. The main results of the study are extremely 
    low AODs during winter, a remarkable increase to high values in summer and a distinct change in the spectral 
    dependencies of AODs from a relatively steeper spectra during winter to a shallower one in summer. A comparison of 
    the total optical depths of the night-time measurements taken during 1970's with the day-time values from the current 
    study, underlines the fact 
    that at those times also the loading of larger size particle occurred during summer though less severely as of today.
    During transparent days, the AOD values lie usually below 0.08 while during dusty (turbid) days, it lies between 
    0.08 to 0.69 at 0.5 $\mu$m. The average AOD value at 0.5 $\mu$m during winters, particularly in January and February, 
    is $\sim 0.03\pm0.01$.
    The mean aerosol extinction law at Manora Peak during 2002 is best represented by $0.10 \lambda^{-0.61}$. However during 
    transparent days, which almost covers 40\% of the time, it is represented by $0.02 \lambda^{-0.97}$. This value of wavelength 
    exponent,
    representing reduced coarse concentration and presence of fine aerosols, indicates that the station measures aerosol in the 
    free troposphere at least during part of the year.

\end{abstract}

\begin{article}

\section{Introduction}
Aerosols, both natural and anthropogenic, play an important role in atmospheric as well as astronomical sciences. In the 
former, they affect by imparting radiative forcing and perturbing the radiative balance of the Earth-atmosphere system as 
well as by degrading the environment. To understand the effects of aerosols on our geo/biosphere systems, it is essential 
to characterize their physical, chemical and optical properties at as many locations as possible because of the regional 
nature of their properties and the short lifetime (Moorthy et al. 1999; Satheesh et al. 2002). This will also help 
in building up a comprehensive picture of global aerosol distribution and also their potential environmental impacts. As 
most of the aerosol sources are of terrestrial origin the variability of their properties will be very large close to the 
surface. At higher altitudes above the mixing region and in the free troposphere, the aerosol characteristics have a more 
synoptic perspective; would be indicative of the background level and are useful to understand long-term impacts. Such 
systematic measurements of aerosols at high altitudes are practically non-existent in India.

For ground-based optical astronomical observations also, precise knowledge of the Earth's atmospheric extinction behaviour
above a site is essential. Amongst many factors which cause extinction of light, the one due to scattering by aerosols is 
highly variable and controls the transparency as well as stability of the sky. Therefore characterisation of atmospheric 
extinction at a site takes special importance in different atmospheric conditions like polluted or clear sky. Studying 
behaviour of aerosol variation does help in evolving an average extinction law as well as in dictating the quality of 
the site. Though some studies on the night-time spectral behaviour of aerosol optical depths (AODs) were done earlier
(Kumar et al. 2000), a multi-wavelength study with narrow band filters have not been undertaken so far.

Realising the potential and need of aerosol studies from both astronomical and atmospheric science perspectives, a programme
has been initiated at Manora Peak, Nainital as a collaborative activity between the Space Physics Laboratory (SPL), 
Thiruvananthapuram and the State Observatory, Nainital. Preliminary results,  based on observations obtained during January 
to June 2002, are presented by Sagar et al. (2002). In this paper, we present the results of extensive measurements of 
spectral AODs and deduced aerosol characteristics based on one year (January to December 2002) observational data. The 
results are discussed and compared with similar measurements obtained during night-time in the 1970's.

\section{Experimental site, observational data and analysis}
The experimental site, Manora Peak, just south-west of Nainital, headquarters of the State Observatory, is located in the 
Shivalik ranges of central Himalayas (latitude $=$ 29$^{0}$22$^{\prime}$ N, longitude $=$ 79$^{0}$27$^{\prime}$ E) at an 
altitude of $\sim$ 2 km. The geographical location of the site over the Indian subcontinent is given in Fig. 1. The day-time 
observations at Manora Peak on aerosols were made for 163 days during January to 
December 2002 with a Multi-Wavelength Solar Radiometer (MWR) designed and developed by SPL. The instrument provides 
columnar total optical depths (TODs) at ten narrow band wavelengths (FWHM of 6 to 10 nm, at different wavelengths) centered at 
0.38, 0.40, 0.45, 0.50, 0.60, 0.65, 0.75, 0.85, 0.935 and 1.025 $\mu$m by making continuous spectral extinction measurements 
of directly transmitted solar radiation. The instrument works on the principle of filter wheel radiometer (Shaw et al. 1973)
and has a field of view $< 2^{0}$.
More details of the instrument and the principle of data reduction and error budget are described by Moorthy et al. (1999, 
2001). Manora Peak being a high altitude station and experiencing very low AODs for a considerable duration of the year, 
meets the requirement for accurate calibration of the Sun-photometer (Shaw 1976).

The observed flux, $F(\lambda, z)$, of the Sun at zenith distance, z, suffers extinction due to the Earth's atmosphere.
It is proportional to \begin{math} e^{-\tau_{\lambda} M(z)} \end{math}, where $\tau_{\lambda}$ is the columnar TOD of the Earth's 
atmosphere at wavelength $\lambda$ and $M(z)$ is the relative airmass at z and is estimated following the general expression 
given by Kondratyev (1969). As the output $V_{\lambda}$ of the MWR at any wavelength is directly proportional to $F(\lambda, z)$,
a least squares linear fit between the natural logarithm of $V_{\lambda}$ and the corresponding relative airmass,
$\tau_{\lambda}$ is used to estimate following the so called 'Langley plot' method (Shaw 1976). Contributions towards 
$\tau_{\lambda}$ are due to scattering by air molecules ($\tau_{R \lambda}$), aerosols ($\tau_{p \lambda}$) and due to 
gaseous absorption ($\tau_{\alpha \lambda}$). The $\tau_{\alpha \lambda}$ consists of molecular absorption due to ozone 
($\tau_{oz \lambda}$), water vapour ($\tau_{w \lambda}$) and other gases such as NO$_{2}$. The $\tau_{R\lambda}$ values are 
estimated analytically using model/reference atmosphere profiles given by Sasi and Sen Gupta (1979). They are 0.361, 0.291, 0.179, 
0.116, 0.055, 0.040, 0.022, 0.014, 0.009 and 0.006 respectively at the wavelengths under consideration. The ozone absorption 
mainly contributes in the Chappuis bands centered at 0.575 $\mu$m (Guti\'{e}rrez-Moreno et al. 1982). The significant values 
of $\tau_{oz \lambda}$ are 0.008, 0.032 and 0.017 at $\lambda = $ 0.5, 0.6 and 0.65 $\mu$m respectively and zero at other 
wavelengths. There is also a weak absorption by NO$_{2}$ at wavelengths below 0.45 $\mu$m. The typical value of optical depth due 
to this is $\sim$ 0.006 (Tomasi et al. 1985).  The $\tau_{w \lambda}$ at affected wavelengths is determined following 
Nair and Moorthy (1998) using MWR measurements at 0.935 $\mu$m. 

\begin{figure}[t]
\noindent\includegraphics[width=18pc]{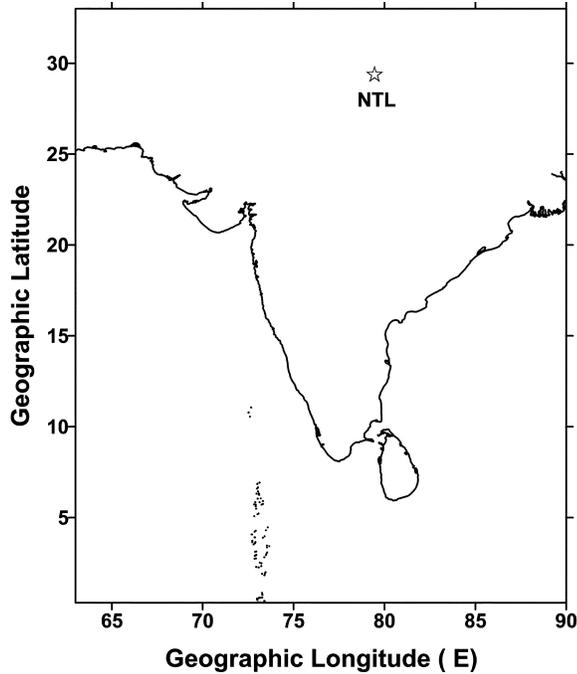}
\caption {The geographical location of Manora Peak, Nainital (NTL) over Indian subcontinent is given.}
\end{figure}

\subsection{Day-time MWR measurements}
The MWR data obtained during the study period have been analysed as described above. On certain days, the Langley plots 
revealed occurrence of different least square linear fit to the data with distinctly different slopes for the forenoon (FN) and 
afternoon (AN) parts of the same day implying different $\tau_{\lambda}$. The  MWR data are therefore analysed separately 
considering the FN and AN part of the data as two independent sets. A consolidated log of the data thus analysed is presented
in Table 1. In all, 241 data sets, obtained on 163 days of the year 2002 have been analysed. The data sets are spread over all 
months of the year except for July and August with maximum of 47 and minimum of 5 data sets for the months of November and 
September 2002 respectively (see Table 1). The complete absence of data in July and August and fewer observations in June 
and September are due to the highly cloudy sky conditions prevailing over the site, associated with the Indian summer monsoon
(Asnani 1993). The values of $\tau_{p \lambda}$ (= $\tau_{\lambda}$ - $\tau_{R\lambda}$ -$\tau_{\alpha \lambda}$) at the 
wavelengths under study for each data set are given in Table 2 in electronic form. Filters at $\lambda = $ 0.40 and 0.60 $\mu$m 
could be installed only on 11 October 2002 and 10 December 2002 
respectively. The number of data sets at these wavelengths are therefore less. However, they will not affect the conclusions
drawn in this paper. 

In order to check the stability and the quality of data obtained from MWR instrument, we study intercepts of all the 
Langley plots (i.e. ln(V$_{\lambda}$) values at zero airmass) against time for filters at 0.38, 0.5 and 0.75 $\mu$m.
The distribution of the corresponding data points are best represented respectively by mean $\pm$ s.d. of $1.18\pm0.08$, 
$1.20\pm0.06$ and $0.78\pm0.05$. The overall errors in the measurements of $\tau_{p\lambda}$ at any wavelength are given as 
\begin{math} \sigma_{\tau_{p\lambda }}^{2} = \sigma_{\tau_{\lambda} }^{2} + \sigma_{\tau_{R \lambda} }^{2} + 
\sigma_{\tau_{\alpha \lambda} }^{2} \end{math}. The $\sigma_{\tau_{\lambda}}$ arises due to 1-sec resolution in time for 
airmass calculation, the statistical errors in regression analysis and the errors due to variation in the Langley intercepts.
All these added together contributes to $<$ 0.02. The error in $\sigma_{\tau_{R \lambda}}$ arises due to atmosphere models 
considered and it may vary by 1\%. The $\sigma_{\tau_{R \lambda}}$ values will therefore be $<$ 0.01. Ozone models also vary 
which may contribute an uncertainty of 0.003 at wavelengths between 0.5 to 0.65 $\mu$m. The AODs may therefore have a maximum 
uncertainty of $\sim$ 0.03 at the wavelengths under considerations.

\begin{table}[t]
\caption{Observing log of the day-time aerosol measurements.}

\begin{tabular}{c|c|c|c|c}
\hline
 Months    & Days & \multicolumn{3}{c}{Data sets} \\ \cline{3-5} 
 (2002)    &      &    Forenoon & Afternoon & Total \\ \hline
 Jan   & 12   & 11   & 06   & 17    \\
 Feb   & 11   & 11   & 02   & 13    \\
 Mar   & 21   & 21   & 16   & 37    \\
 Apr   & 21   & 19   & 13   & 32    \\
 May   & 10   &  9   & 03   & 12    \\
 Jun   & 07   & 06   & 02   & 08    \\ 
 Sep   & 05   & 05   & 00   & 05    \\ 
 Oct   & 25   & 25   & 10   & 35    \\
 Nov   & 27   & 27   & 20   & 47    \\
 Dec   & 24   & 19   & 16   & 35    \\  \hline
Total      &163   &153   & 88   &241    \\ \hline
\end{tabular}
\end{table}

\subsection{Night-time measurements}
As a part of optical astronomical observations, the night-time values of $\tau_{\lambda}$ were determined on 14 nights
during 1970 to 1978 at 20 wavelengths ranging from 0.34 $\mu$m to 0.76 $\mu$m with a bandwidth of 5 nm (details are given in 
electronic form in Table 3). The observations were obtained using 0.5 and 1 meter size optical telescopes along with a star 
photometer. Further details of the observations and measurements are given by Kumar et al. (2000). The $\tau_{\lambda}$ 
measurements can therefore be considered monochromatic and hence similar to the
MWR observations. The accuracy of the night-time measurements lies between 0.01 and 0.02 airmass$^{-1}$ being maximum 
shortward of 0.4167 $\mu$m. It is thus similar to that of the day-time measurements. They can therefore be compared with 
the day-time MWR measurements for the study of spectral properties but not for the study of monthly variations due to the 
rather short database.

\section{Results and discussions}

\begin{table*}[t]
{{\bf Table 5.} Monthly mean AODs (at 0.5 $\mu$m), water vapour content (W) and aerosol wavelength exponent ($\alpha$)} 
\smallskip
\begin{tabular}{c|c|c|c|c|c|c|c|c|c}  \hline
Months&\multicolumn{3}{c|}{AODs at 0.5$\mu$m}&\multicolumn{3}{c|}{W (g cm$^{-2}$)}&\multicolumn{3}{c}{$\alpha$}\\\cline{2-10} 
(2002)     & Min& Mean$\pm$sd& Max&Min& Mean$\pm$sd& Max&Min& Mean$\pm$sd& Max \\ \hline
Jan&  0.00& 0.04$\pm$ 0.01& 0.11&  0.00&  0.06$\pm$ 0.01&  0.16&  1.55&  1.90$\pm$0.27& 2.61  \\
Feb&  0.01& 0.03$\pm$ 0.01& 0.09&  0.03&  0.07$\pm$ 0.01&  0.11&  0.78&  1.52$\pm$0.18& 2.52  \\ 
Mar& -0.01& 0.16$\pm$ 0.02& 0.44&  0.02&  0.18$\pm$ 0.02&  0.40&  0.17&  1.08$\pm$0.09& 1.76  \\
Apr&  0.06& 0.29$\pm$ 0.03& 0.67&  0.06&  0.24$\pm$ 0.02&  0.72&  0.01&  0.70$\pm$0.16& 1.30  \\ 
May&  0.05& 0.32$\pm$ 0.04& 0.56&  0.05&  0.19$\pm$ 0.02&  0.28&  0.42&  0.68$\pm$0.16& 1.35  \\
Jun&  0.22& 0.36$\pm$ 0.05& 0.69&  0.19&  0.46$\pm$ 0.06&  0.74&  0.21&  0.47$\pm$0.52& 0.91  \\ 
Sep&  0.05& 0.06$\pm$ 0.01& 0.10&  0.06&  0.13$\pm$ 0.04&  0.29&  0.40&  1.08$\pm$0.26& 1.46  \\
Oct&  0.03& 0.11$\pm$ 0.01& 0.37&  0.04&  0.15$\pm$ 0.01&  0.32&  0.37&  1.20$\pm$0.11& 2.22  \\ 
Nov& -0.02& 0.08$\pm$ 0.01& 0.33&  0.02&  0.09$\pm$ 0.01&  0.22&  0.40&  1.45$\pm$0.07& 2.81  \\
Dec& -0.01& 0.12$\pm$ 0.02& 0.50&  0.00&  0.06$\pm$ 0.01&  0.18&  0.10&  1.26$\pm$0.13& 2.76  \\  \hline
\end{tabular}
\end{table*}

\subsection{Temporal variation of AODs}
Fig. 2 shows the temporal variations of AODs at the representative wavelength, 0.5 $\mu$m and water vapour content in panels 
(a) and (b) respectively. During 15 June to 15 September no AODs observations could be taken due reasons discussed above. 
The data cover the range from excellent "coronal" days where $\tau_{p \lambda}$ is very low ($<$ 0.1) to 
"absorbent" (with high $\tau_{p \lambda}$ $>$ 0.4 ) days where summer dust from the adjoining plain areas is present above 
the site. Fortnightly or the monthly means along with minimum and maximum AOD values at all the MWR wavelengths during the 
study period are given in Table 4, again in electronic form due to its large size.
In order to have statistically significant results, monthly means are taken if the number of data sets in a month are $<$ 20.
A plot of these at 0.5 $\mu$m are also shown by the continuous lines in Fig. 2. The upper panels of Fig. 2 show the mean value 
of columnar water vapour content, W (g/cm$^2$), 
relative humidity (\%), rainfall (cm), wind speed (m/s) and wind direction (deg). These plots clearly show correlated 
variations. The monthly mean values of AOD at 0.5 $\mu$m and W, along with the respective standard deviations and ranges are given 
in Table 5. The average values of $\tau_{p \lambda}$  at 0.50 $\mu$m are $\sim 0.03\pm0.01$ during 
January and February, between 0.06 to 0.11 during September to 15 December and in the range of 0.16 to 0.36 during 15 March 
to June. The AOD values are thus lowest during winter and post-monsoon (January to mid-March and October to December). 
However, it increases rather rapidly attaining a peak value during summer (April to June) following a transition in the month
of March. The next transition from peak to low, might be occurring during the rainy months when the MWR data are absent.

The monthly variation of major surface meteorological parameters at the site are also shown in Fig. 2 to study its possible 
effects on aerosols. The water vapour content, W, is below 0.2 g/cm$^2$ for 80\% of the measurements with a mean 
value of 0.1 g/cm$^2$ implying a dry environment, while it is greater than 0.2 g/cm$^2$ mostly in summer and in the month of 
October. An increase in W 
(when  higher than 3 g/cm$^2$ ) is known to cause a non-linear increase in AOD at all wavelengths covered by MWR 
(Nair and Moorthy 1998; Moorthy et al. 2001). As the values of W are very low at the site, the temporal variations in W and 
AOD are quite similar (see Fig. 2), with the only exception for the month of October, where AODs are small despite large 
($>$ 0.2 g/cm$^2$) values of W. This indicates that the variations in AOD and W are brought about mostly by the same processes 
related to local and synoptic changes in the meteorological conditions. The rainfall climatology shows that the rainfall is 
highest during July to September (accounting for about 68\% of the annual) with very little rain during the period March to 
middle of June (when it is only 12\% of the annual). The top two panels of Fig. 2 give the mean wind speed and wind direction
of arrival 
recorded during 2002 at the site. The arrival wind direction is indicated in degrees with 0$^{0}$, 90$^{0}$, 180$^{0}$ and 
270$^{0}$ corresponding to North, East, South and West directions respectively. The mean wind speed, $\sim$ 2 m s$^{-1}$ during 
winter,
doubles during pre-monsoon. During the same period the mean arrival wind direction gradually shifts from Southerly to Westerly.
For the location of the site this would mean a shift in the airmass from the Southern (with respect to the station) Indian plains 
in winter to the Western arid land mass during summer, when the winds arrive from the vast arid regions of North-West India and 
regions lying further 
to its West. This change in the airmass type, obviously is mostly responsible for the rapid build-up in the AOD over the station 
after March as the arid airmass is known to transport large amounts of desert/mineral aerosols from the West Asian and Indian 
deserts, lying to the West/South-West of this station. The role of wind blown dust from the deserts in increasing turbidity over 
Northern Indian regions have been suggested earlier by Mani et al. (1969). Recently, based on satellite data analysis, 
Li and Ramanathan 
(2002) have also shown the Eastward transport of West-Asian deserts aerosols across the Northern Arabian sea towards the West 
coast of India. Thus the observed sharp increase in the AOD from middle of March is mostly attributed to this. In addition to 
the advection by airmass, the increased solar heating of the land mass over the lower plains adjacent to the site during the 
summer season would results in increased convective mixing and elevation of the boundary layer aerosols. This would also 
contribute to the increase in AOD over the site during the summer season.

\begin{figure}[t]
\noindent\includegraphics[width=20pc]{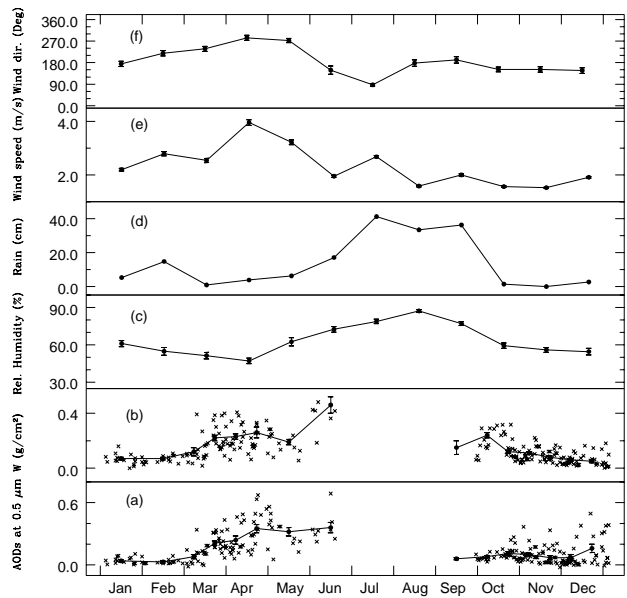}
\caption {Temporal variations of the day-time AOD at 0.5 $\mu$m during the year 2002 
are shown in panel (a). The panel (b) shows the corresponding variations in water vapour content. The fortnightly/monthly 
mean are also plotted in these panels (see text). Monthly variations of relative humidity, rainfall, wind speed and mean 
arrival wind direction are shown in panels (c), (d), (e) and (f) respectively.}
\end{figure}

\begin{figure}[t]
\noindent\includegraphics[width=20pc]{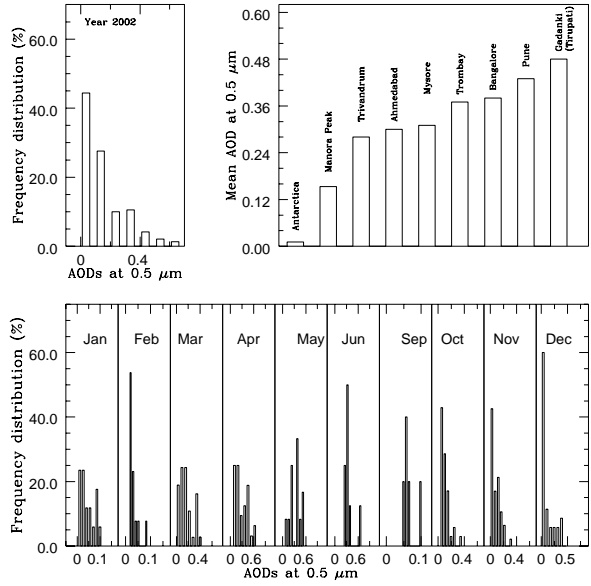}
\caption {Lower panels show monthly variation of frequency distribution of AODs at 0.5 $\mu$m during the year 2002, while the 
upper panels show its yearly frequency distribution and comparison with those at Antarctic and other Indian locations.}
\end{figure}

\begin{figure}[t]
\noindent\includegraphics[width=20pc]{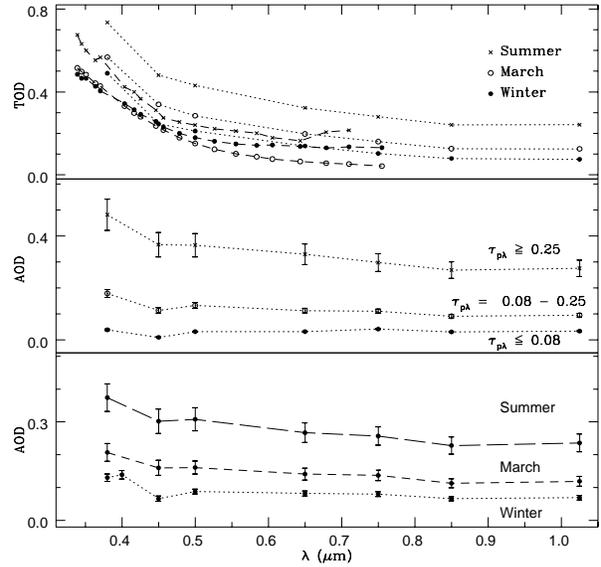}
\caption {Lower panel shows spectral variation of day-time AODs during different seasons while the middle panel displays
the spectral variation of different $\tau_{p \lambda}$ at 0.5 $\mu$m. Upper panel compares the day as well as night-time
total optical depths, with dotted and dashed lines respectively.}
\end{figure}

\begin{figure}[t]
\noindent\includegraphics[width=20pc]{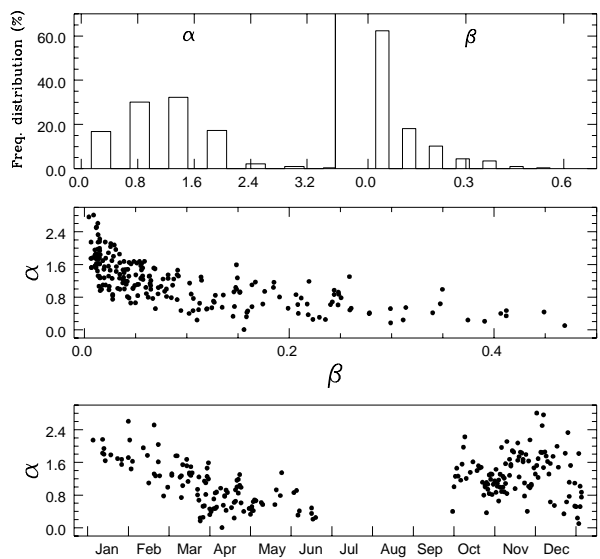}
\caption {Variation of $\alpha$ with time is shown in the lower panel.
 A correlation between $\alpha$ and $\beta$ and there frequency distributions
over the entire period are presented in the other panels.}
\end{figure}

The AOD values for each month as well as for the entire year are grouped together and the percentage frequency distributions
for different $\tau_{p \lambda}$ ranges have been plotted in Fig. 3. Here again, it is observed that day-to-day variability 
is maximum during April, May and June. The distributions are highly skewed in winter months, while in summer they tend 
to be more symmetric. These distributions support the conclusions drawn earlier.

\subsection{Comparison of AODs with other locations}
The yearly mean values of AODs at the site are 0.180, 0.174, 0.163, 0.153, 0.137, 0.130, 0.119, 0.110, 0.104 and 0.098 
respectively for the ten wavelengths under investigation. However, it may be noted that the corresponding mean AOD values 
are 0.043, 0.041, 0.036, 0.033, 0.028, 0.026, 0.022, 0.020, 0.018 and 0.017 respectively during winter. 
The yearly mean $\tau_{p\lambda}$ values over Manora Peak, representative of 10 months period during 2002, are higher to 
that observed over Antarctica during 3 months of summer period (Gadhavi and Jayaraman 2002), while the winter values are 
similar to the values recorded at Antarctic environment (Fig. 3). This indicates a very clean environment over Manora Peak 
at least during part of the year 2002. 
 The yearly mean values of  $\tau_{p\lambda}$ for Manora Peak are much lower than the average values observed
over typical urban regions like Ahmedabad (Ganguly et al. 2002), Trombay (Sunny et al. 2002), Bangalore (Babu et al. 2002),
Gadanki (Rao et al. 2002), Pune (Pandithurai et al. 2002) during the year 2002 (see Fig. 3). Based on over ten years of 
observations, Moorthy et al. (1999) have reported annual mean values of AOD at 0.5 $\mu$m as 0.28 for Trivandrum; 0.31 for 
Mysore; and 0.5 for Gadanki.

\subsection{Spectral variation of AODs}
The seasonal variations of the AOD spectral properties are plotted in Fig. 4 (lower panel), grouped in winter, summer and 
March (transition month as discussed earlier). The AODs are averaged during these periods. The vertical bars show the standard 
deviations. 
The properties of winter and summer aerosols are different as can be seen from the variations of their $\lambda$ dependence. 

To study the spectral differences as a function of $\tau_{p \lambda}$, we grouped the entire data in three groups according
to $\tau_{p \lambda}$ values at 0.5 $\mu$m. The groups are $\tau_{p \lambda}$ $<$ 0.08, 0.08 $\le$ $\tau_{p \lambda}$ $<$ 0.25 and
$\tau_{p \lambda}$ $\ge$ 0.25. The wavelength dependence of AOD for these groups are displayed in the middle panel of Fig. 4.
This clearly indicates the differences between the size of particles contributing to the AODs of different groups. The
spectral variation of the group with largest values of AOD correspond to summer period, as expected. However, the spectral
variation of the group with lowest AODs is not similar to the winter period.

The upper panel of Fig. 4 shows a comparison of TODs observed for the year 2002 with those measured 
during night-time in seventies. In order to avoid crowding, the  standard deviations are not plotted. The number of data sets 
are less for the night-time measurements. Nevertheless, the summer-winter trend is apparent even in the night-time in seventies,
however the differences between the two values are not as high as it is for the year 2002 during day-time. This would be due to 
several reasons. It is possible that the general increase in AOD at some Indian stations reported by Moorthy et al. (1999) and 
Satheesh et al. (2002) might be contributing to an increase in AOD in the free troposphere, due to increased aerosol loading. 
Particularly in summer, another possibility is that the increased convective turbulence during pre-monsoon/summer months would 
be pumping in more surface
level aerosols to higher altitudes, which would decrease in the night-time due to the collapse in the boundary layer height.
This would lead to higher AODs in day-time. This effect could be more conspicuous at short wavelengths as small particles can 
easily be lifted by turbulence. This aspect can be verified only by making both night-time and day-time measurements during 
the same period.

An interesting feature of the TODs, common to both the day-time and night-time measurements, is its ultra-low values for a 
few percent of the total measurements. The occurrence of these values are more prominent for the night-time  measurements 
during 1970's indicating a clearer atmosphere (see Fig. 4 top panel). The lowest day-time TOD was obtained for 
the FN dataset of 04 March 2002. The rain poured heavily previous to this day. Similarly the log hints ultra-clear or pure 
skies on these dates. It may be noted that these ultra-low values of TOD are generally obtained during November to March of 
the year between 0.34 to 0.45 $\mu$m and are more pronounced for night-time measurements. At 0.36 
$\mu$m, the TOD values lower by $\sim$ 0.1 than expected from standard atmosphere models have also been obtained at other 
central Himalayan sites, e.g., at Devasthal (latitude $=$ 29$^{0}$22$^{\prime}$ N, longitude $=$ 79$^{0}$41$^{\prime}$ E, 
altitude $=$ 2450 m) by Mohan et al. (1999) and at Hanle (latitude $=$ 32$^{0}$47$^{\prime}$ N, longitude $=$ 78$^{0}$58$^{\prime}$
E, altitude $\sim$ 4500 m) by Parihar et al. (2003) during night-time measurements using broad band (half width $\sim$ 200 \AA)
filters. The occurrence of negative AODs mostly at shorter wavelengths may arise either due to changes in the climatological values
of ozone column content at the central Himalayan sites or the use of model atmosphere as that obtained from a tropical station 
by Sasi and Sen Gupta (1979). Thus overestimation of Rayleigh and ozone corrections may result in the negative AODs. It could 
also be due to very high value of Rayleigh optical depth ($\sim$ 90\% of the TODs at these wavelengths) being subtracted from 
$\tau_{\lambda}$. 
Also the uncertainty of $\pm 0.03$ in AOD determination does cover most of these negative values, however, these values may be 
used to constrain the existing models of standard Earth atmosphere for the Himalayan region.

\subsection{Determination  of aerosol properties / Angstr\"{o}m coefficients}
Following Angstr\"{o}m (1961), the wavelength variation of AOD can be expressed as \begin{math} 
\beta \lambda^{-\alpha} \end{math}, where $\beta$ and $\alpha$ are the constants and vary widely. The wavelength exponent 
$\alpha$ is a measure of the relative dominance of fine, sub-micrometer sized aerosols over the coarse aerosols while 
$\beta$ is a measure of the total aerosol loading. Higher value of $\alpha$ signifies increased relative abundance of fine 
particles. The values of both $\alpha$ and $\beta$ are determined from the individual data sets of the present MWR observations.
The correlation of $\alpha$ with $\beta$ is shown in the middle panel of Fig. 5. We see that as $\beta$ increases, $\alpha$ 
decreases (though the relation is not linear). This indicates that as the aerosol loading increases, the relative dominance 
of fine aerosol decreases. In other words, the increase in aerosol loading is more due to increase in coarse aerosol particles.

Temporal variation of $\alpha$ is shown in Fig. 5 along with the frequency distributions of both $\alpha$ and $\beta$.
The monthly mean values of $\alpha$ along with the range are given in the Table 5.
The average value of $\alpha$ decreases systematically from January to March and remains at that level till June, while it shows
large scatter during October to December with increased average. Histograms indicate highly skewed distributions for $\beta$
with most of the values $<$ 0.1 while $\alpha$ values have a wide Gaussian distribution with peak around 1 
(see Fig. 5 upper most panel).

We have also derived the values of $\alpha$ and $\beta$ by fitting the aerosol extinction law in the mean values of AODs for 
different range of AODs, seasons as well as for the whole year. The values of $\beta$ and $\alpha$ are $0.017\pm0.005$ and
$0.97\pm0.09$ when AOD at 0.5 $\mu$m $<$ 0.08, while the corresponding values for AOD $\ge$ 0.08 are $0.153\pm0.005$ and  
$0.54\pm0.07$ respectively. The values of $\beta$ for the winter, March and summer are $0.06\pm0.01$, $0.11\pm0.01$ and 
$0.22\pm0.01$ respectively, while the corresponding values of $\alpha$ are $0.72\pm0.12$, $0.54\pm0.08$ and 
$0.46\pm0.07$ respectively. The decrease in $\alpha$ signifies increase in the relative abundance of coarse mode aerosols. For 
the entire data set, mean values of $\beta$ and $\alpha$ are found to be $0.10\pm0.01$ and 
$0.61\pm0.08$ respectively. The mean values of $\alpha$ decreases from winter to summer while during the same period, 
$\beta$ value is enhanced by a factor of 4. This observation is important for aerosol characterisation at the site. During 
winter (when land temperatures are very low (0 to 19$^{0}$ C) and minimum solar zenith angle is $\geq$ 40$^{0}$) the surface 
convective
(thermal) activities will be very weak. The observation site being at $\sim$ 2 km above MSL, well above the mixed layer, its 
environment will thus be free of all local contaminations and aerosol characteristics will be pertaining more close to that of 
the free troposphere. In this region, the dominating aerosols will be sub-micron sized ones formed either in situ by secondary 
gas to particle conversion processes of the precursors (which might be more of anthropogenic and regional in nature). These 
fine aerosols would rapidly undergo size transformation by coagulation and condensation growth to accumulation size range.
The aerosol size spectrum would thus be dominating by these particles and hence the AOD spectra during winter would be steeper 
with higher value of $\alpha$ and low value of $\beta$. As Sun enters the northern hemisphere, the surface heating increases 
and there is a better exchange between boundary layer and free troposphere due to increased convective mixing and the increase
in the altitude extent of the convective boundary layer. This is conducive for the local aerosols (which are likely to have a 
large share of coarse particles) to impact the site. In addition to these the arid aerosols, advected by the Westerly winds 
would also contribute to increased coarse particle abundance. As a result the total aerosol loading and the share of coarse 
aerosol to it increases. This is reflected by the steady increase in $\beta$, decrease in $\alpha$ and the flattening of AOD 
spectrum. This continues till the monsoon rains intervene and remove the aerosols by scavenging act.

\section{Conclusions}
The main conclusions of our study are the following :-  
   \begin{enumerate}

   \item AODs over Manora Peak shows significant temporal variations during the year 2002. There is a remarkable increase 
         in AOD from their very low values in winter to high values in summer. During winter season, the AODs are 
         of magnitude comparable to the Antarctic environment, while during summer they are typical to continental regions.

    \item During transparent days the AOD at $0.5 \mu$m lie usually below 0.08 while during dusty days, 
          it lies between 0.08 to 0.69. The mean aerosol extinction law at the site during 2002 is best represented by 
          $0.10 \lambda^{-0.61}$. During transparent days, for 40\% of the observable days, it is $0.02 \lambda^{-0.97}$.

    \item A comparison of the total optical depths of the night-time measurements, taken during 1970's,
          with the day-time underlines the fact that the summer sky pollution occurred at those times also, though less
          severely as of today.

    \item The water vapour content, W, lies $<$ 0.2 g/cm$^2$, for 80\% of the clear days.

   \end{enumerate}

\begin{acknowledgements} 
The authors acknowledge the anonymous reviewers whose useful comments have improved the paper substantially. 
We gratefully acknowledge the initiative and keen interest taken by Prof. R. Sridharan, Director, SPL, in this 
collaborative program. We also thank Dr. Wahab Uddin and  technical staff of the State Observatory, Nainital for providing 
valuable help during observations and to the technical staff of SPL, Trivandrum for installing the instrument at the site.
\end{acknowledgements}

\end{article}


\begin{thebibliography}{}
\bibitem{}Angstr\"{o}m, A., Techniques of determining the turbidity of the atmosphere, {\it Tellus, 13,} 214, 1961.
\bibitem{}Asnani, G.C., Tropical Meteorology Vol. I, Indian Institute of Tropical Meteorology, Pune, India, 1993.
\bibitem{}Babu, S.S., Satheesh, S.K., Moorthy, K.K., \& Vinoj, V., Aerosol radiative forcing over land due to black carbon
          aerosols, {\it Bull. of Indian Aerosol Science and Technology Association,  14,} 88, 2002. 
\bibitem{}Gadhavi, H., \& Jayaraman, A., Direct aerosol radiative forcing experiment over antarctica, 
          {\it Bull. of Indian Aerosol Science and Technology Association, 14,} 40, 2002.
\bibitem{}Ganguly, D., Gadhavi, H., \& Jayaraman, A., Pre-monsoon aerosol characteristics over Ahmedabad,  
          {\it Bull. of Indian Aerosol Science and Technology Association, 14,} 37, 2002.
\bibitem{}Guti\'{e}rrez-Moreno, A., Moreno, H., \& Cort\'{e}s, G.,  A study of atmospheric extinction at Cerro Tololo 
          Inter-American observatory, {\it PASP, 94,} 722, 1982.
\bibitem{}Kondratyev, K.Ya., 1969, Radiation in the atmosphere, Academic press, New York
\bibitem{}Kumar, B., Sagar, R., Rautela, B.S., Srivastava, J.B., \& Srivastava, R.K.,  Sky transparency over 
          Nainital: a retrospective study, {\it Bull. Astron. Soc. India,  28,} 675, 2000.
\bibitem{}Li, F., \& Ramanathan, V., A winter to summer monsoon variation of aerosol optical depth over the tropical Indian 
          ocean, {\it J. Geophys. Res., 107,} 4284, 2002. 
\bibitem{}Mani, A., Chacko, O., \& Hariharn, S., A study of Angstr\"{o}m's turbidity parameters from solar radiation 
          measurements in India, {\it Tellus, 21,} 829, 1969.
\bibitem{}Mohan, V., Uddin, W., Sagar, R., \& Gupta, S. K., Atmospheric extinction at Devasthal, Nainital, 
          {\it Bull. Astron. Soc. India,  27,} 601, 1999.
\bibitem{}Moorthy, K.K., Niranjan, K., Narasimhamurthy, B., Agashe, V.V., \& Murthy, B.V.K., Aerosol Climatology over 
         India, 1 - ISRO GBP MWR Network and Database, ISRO GBP SR-03-99, 1999.
\bibitem{}Moorthy, K.K., Saha, A.., Prasad,B.S.N., Niranjan, K., Jhurry, D., \& Pillai, P.S., Aerosol optical depths over 
          peninsular India and adjoining oceans during the INDOEX campaigns: Spatial, temporal, and spectral characteristics,
          {\it J. Geophys. Res. 106,} 28539, 2001. 
\bibitem{}Nair, P.R., \& Moorthy K.K., Effects of changes in the atmospheric water vapour content on the physical properties
          of atmospheric aerosols at a coastal station, {\it J. Atmos. Solar-Terr. Phys., 60,} 563, 1998.
\bibitem{}Pandithurai, G., Pinker, R.T., \& Devara, P.C.S., Variability of climatically important aerosol optical properties
          over an urban tropical site as retrieved from sky radiometer observation, {\it Bull. of Indian Aerosol Science and 
          Technology Association, 14,} 19, 2002.
\bibitem{}Parihar, P.S., Sahu, D.K., Bhatt, B.C., Subramaniam, A., Anupama, G.C., Prabhu, T.P., 
          {\it Bull. Astron. Soc. India,} in press, 2003. 
\bibitem{}Rao, Y.J., Devara, P.C.S., Srivastava, A.K., Sonbawne, S., \& Kumar, Y.B., LIDAR and radiometric observations
          of aerosols over Gadanki (13.8N, 79.2E), {\it Bull. of Indian Aerosol Science and Technology Association, 14,} 23, 2002. 
\bibitem{}Sagar, R., Kumar, B., Pant, P., Dumka, U.C., Moorthy, K.K., \& Sridharan, R., Aerosol contents at an altitude
          of ~2km in central Himalayas, {\it Bull. of Indian Aerosol Science and Technology Association, 14,} 
          167, 2002; physics/0211040
\bibitem{}Sasi, M.N., \& Sen Gupta, K, Sci. Rep. ISRO, VSSC, SR,  19, Vikram Sarabhai Space Centre, Trivandrum, 72, 1979.
\bibitem{}Satheesh, S.K., Ramanathan, V., Holben, B.N., Moorthy, K.K., Loeb N.G., Maring H., Prospero J.M., \& Savoie D.,
          Chemical, microphysical, and radiative effects of Indian Ocean aerosols, {\it J. Geophys. Res., 107,} 4725, 2002.
\bibitem{}Shaw, G.E.,  Error analysis of multiwavelength Sun photometry, {\it Pure Appl. Geophys., 114,} 1, 1976.
\bibitem{}Shaw, G.E., Regan, J.A., \& Herman, B.M., Investigations of atmospheric extinctions using direct solar radiation
          measurements made with a multiple wavelength radiometer, {\it J. Appl. Meteoro.,  12,} 374, 1973.
\bibitem{}Sunny, F., Indumati, S., \& Daoo, V.J., Preliminary measurements of aerosol optical thickness, columnar content of
          ozone and precipitable water at Trombay, {\it Bull. of Indian Aerosol Science and Technology Association, 14,} 91, 2002.
\bibitem{}Tomasi, C., Marani, S., \& Vitale, V., Multiwavelength Sunphotometer calibration corrected on the basis of the 
          spectral features characterising particulate extinction and nitrogen dioxide absorption, 
          {\it Appl. Opt.,  24,} 2962, 1985.

\end{thebibliography}
\end{document}